\documentclass[pra,eqsecnum,twocolumn, showpacs]{revtex4}
\usepackage{stmaryrd}
\usepackage{amsmath}
\usepackage{amssymb}
\usepackage{graphicx}
\usepackage{textcomp}
\usepackage{calrsfs}
\usepackage{yfonts}

\newcommand{\msum}{\sum_{m=0}^\infty \!{}^{{}^\prime}}

\newcommand{\kp}{k_\perp}
\newcommand{\kz}{k_z}

\newcommand{\be}{\begin{equation}}
\newcommand{\ee}{\end{equation}}
\newcommand{\ben}{\begin{equation*}}
\newcommand{\een}{\end{equation*}}
\newcommand{\bea}{\begin{eqnarray}}
\newcommand{\eea}{\end{eqnarray}}

\newcommand{\br}{\mathbf{r}}

\newcommand{\te}{\tilde {e}}

\newcommand{\gE}{\mbox{\textgoth{E}}}
\newcommand{\tgE}{\tilde{\mbox{\textgoth{E}}}}

\newcommand{\im}{\Im\mathrm{m}}
\newcommand{\re}{\Re\mathrm{e}}

\newcommand{\matD}{\mathbf{D}}
\newcommand{\mattD}{\tilde{\mathbf{D}}}

\newcommand{\Hn}{H_\nu^{(1)}}
\newcommand{\Hnp}{{H_\nu^{(1)}}'}
\newcommand{\Hna}{H_\nu^{(1)}(\kp\rho)}

\newcommand{\Kn}{K_{\nu}}
\newcommand{\Jn}{J_\nu}

\newcommand{\tmE}{\tilde{\mathcal{E}}}
\newcommand{\mE}{\mathcal{E}}

\newcommand{\oa}{\bar{a}}
\newcommand{\ob}{\bar{b}}
\newcommand{\ua}{\underline{a}}
\newcommand{\ub}{\underline{b}}

\newcommand{\Arkw}{(\boldsymbol\rho; k_z,\omega)}
\newcommand{\Antkw}{(\theta;\nu,k_z,\omega)}

\begin{document}

\title{Electrodynamic Casimir Effect in a Medium-Filled Wedge II}

\date{\today}
\author{Simen {\AA}dn{\o}y \surname{Ellingsen}}
\email{simen.a.ellingsen@ntnu.no}
\author{Iver \surname{Brevik}}\email{iver.h.brevik@ntnu.no}
\affiliation{Department of Energy and Process Engineering, Norwegian
University of Science and Technology, N-7491 Trondheim, Norway}
\author{Kimball A. Milton}\email{milton@nhn.ou.edu}
\affiliation{Oklahoma Center for High Energy Physics and Department of
Physics and Astronomy, The University of Oklahoma, Norman, OK 73019, USA}

\begin{abstract}
  We consider the Casimir energy in a geometry of an infinite
magnetodielectric wedge closed by a circularly cylindrical, perfectly reflecting arc embedded in
another magnetodielectric medium, under the condition that the speed of
light be the same in both media. An expression for the Casimir energy
corresponding to the arc is obtained and it is found that in the limit
where the reflectivity of the wedge boundaries tends to unity the finite part of
the Casimir energy of a perfectly conducting wedge-shaped sheet closed by a
circular cylinder is regained. The energy of the latter geometry possesses
divergences due to the presence of sharp corners. We argue how this is a
pathology of the assumption of ideal conductor boundaries, and that no
analogous term enters in the present geometry.
\end{abstract}

\maketitle

The Casimir effect \cite{casimir48} may be understood as an effect of the
fluctuations of the quantum vacuum. Casimir's original geometry involved two
infinite and parallel ideal metal planes which were found to attract each
other with a negative pressure scaling quartically with the inverse interplate
separation. In a seminal paper, Lifshitz generalised Casimir's original
calculation to imperfectly reflecting plates \cite{lifshitz55}. Since its
feeble beginnings research on the Casimir effect has grown from being of
peripheral interest to a few theorists to a bustling field of research both
experimental and theoretical with publications numbering in the hundreds each
year. Recent reviews include
\cite{BookMilton01, milton04, lamoreaux05,buhmann07}.

Progress on Casimir force calculations for other geometries has been slower
in coming. Spherical and cylindrical geometries have naturally been objects
of focus, the latter of direct interest to the effort reported herein. Only
in 1981 was the Casimir energy of an infinitely long perfectly conducting
cylindrical shell calculated \cite{deraad81} and the more physical but also
significantly more involved case of a dielectric cylinder was considered
only in recent years \cite{brevik94, gosdzinsky98,milton99, lambiase99,
caveroPelaez05, romeo05,brevik07}.  We might also mention
recent work on the cylinder defined by a $\delta$-function potential,
a so-called semitransparent cylinder \cite{caveroPelaez06};
for weak-coupling, both the semitransparent cylinder and the dielectric
cylinder have vanishing Casimir energy.

Closely related to the cylindrical geometry is the infinite wedge. The problem
was first approached in the late seventies \cite{dowker78, deutsch79} as part
of the still ongoing debate about how to interpret various divergences in
quantum field theory with sharp boundaries. Since, various embodiments of the
wedge have been treated by Brevik and co-workers \cite{brevik96,brevik98,
brevik01} and others \cite{nesterenko02, razmi05}. A review may be found in
\cite{BookMostepanenko97}. A wedge intercut by a cylindrical shell was
considered by Nesterenko and co-workers, first for a semi-cylinder
\cite{nesterenko01}, then for arbitrary opening angle \cite{nesterenko03},
and the corresponding local stresses were studied by Saharian
\cite{rezaeian02,saharian07, saharian09}. The group at Los Alamos studied the
interaction of an atom with a wedge \cite{mendes08, rosa08} previously
investigated by Barton \cite{barton87} and others \cite{skipsey05, skipsey06},
the geometry realised in an experiment by Sukenik et al.\ some years ago
\cite{sukenik93}. A recent calculation of the Casimir energy of a
magnetodielectric cylinder intercut by a perfectly reflecting wedge filled
with magnetodielectric material was recently reported by the current authors
\cite{brevik09}. Common to all of these theoretical efforts is the assumption
that the wedge be bounded by perfectly conducting walls.

While until recently relatively few treatments of the vacuum energy of the
wedge existed, the problem of calculating the diffraction of electromagnetic
fields by a dielectric wedge within classical eletromagnetics is an old one and
several powerful methods have been developed within this field. The Green's
function of the potential (Poisson) equation in the vicinity of a perfectly
conducting wedge was found more than a century ago by Macdonald
\cite{macdonald1895} and extended to the wave equation with a plane wave source
by Sommerfeld \cite{sommerfeld1896}. Generalising Sommerfeld's method, the
first theoretical solution to the scattering problem involving a wedge of
finite conductivity was found by Malyuzhinets in his PhD work
\cite{malyuzhinetsThesis50} (see \cite{osipov99} for a review; cf.\ also
\cite{osipov98}).

A different method was proposed by Kontorovich and Lebedev in 1938
\cite{kontorovich38} and used by Oberhettinger to solve the Green's function
problem some time later \cite{oberhettinger54}. The method has been given
attention in recent analytical and numerical studies of the diffraction
problem \cite{osipov93, knockaert97, rawlins99, salem06, salem08}.

In the present effort we study the Casimir energy in a magnetodielectric wedge
of opening angle $\alpha$ inside and outside a perfectly conducting cylindrical
shell of radius $a$---See Fig.~\ref{fig_wedge}. The interior and exterior of
the wedge are both filled
with magnetodielectric material under the restriction of isorefractivity (or
diaphanousness), that is, the index of refraction $n^2(\omega)=
\epsilon(\omega)\mu(\omega)$ is the same everywhere for a given frequency.
This condition is adopted because without it the problem is no longer
separable and not readily solvable.  Moreover, we suspect that nondiaphanous
media will lead to divergences, at least in the absence of dispersion.

As a natural extension of the considerations in \cite{brevik09} we derive an
expression for the free energy of such a system by use of the argument
principle \cite{vankampen68}. (By free energy, we mean that bulk terms
not referring to the circular arc boundary are subtracted.)
The necessary dispersion relation provided by the electromagnetic boundary
conditions at the wedge sides is derived in two different ways; by a standard
route of expansion of the solutions in Bessel function partial waves and by use
of the Kantorovich-Lebedev (KL) transform.
(Still a third method, based on the Green's function formulation, is
sketched in the Appendix.) The corresponding boundary condition
equation at the cylindrical shell is well known. These together allow us to sum
the energy of the eigenmodes of the geometry satisfying eigenvalue equations
for the frequency and azimuthal wave number $\nu$ by means of the argument
principle.

There are important differences between the diaphanous geometry considered
herein and the standard geometry of a perfectly conducting wedge. Assuming
diaphanous electromagnetic boundary conditions, the interior and exterior wedge
sectors are coupled and remain so also in the limit where the reflectivity of
the wedge boundaries tends to unity (for example by letting $\epsilon\to\infty,
\mu\to 0$ so that their product is constant). Assuming the wedge be perfectly
conducting from the outset, however, the interior of the wedge is severed
cleanly from its exterior at all frequencies, a significantly different
situation.

The Casimir energy of the perfectly conducting wedge and magnetodielectric arc
considered in \cite{brevik09} was found to possess an unremovable divergent
term associated with the corners where the arc meets the wedge. This is a
typical artifact of quantum field theory with non-flat boundary conditions
(e.g.\ \cite{nesterenko02,nesterenko01,nesterenko03}). We will argue in section
\ref{sec_addterm} that there is no such term present in the geometry considered
herein, and that the direct generalisation of the finite part of the
energy of the system considered in \cite{brevik09} to the present system is in
fact the full regularised Casimir energy. The reason for this rests upon two
unphysical effects of perfectly conducting boundary conditions at the wedge
sides (the vanishing of the tangential components of the electric field
there). Namely, such
boundary conditions exclude the existence of an azimuthally constant TM mode,
and divides space cleanly into an interior and exterior sector with no coupling
allowed between modes in the two sectors. Moreover, for
a wedge consisting of  perfectly conducting thin sheets dividing space into
two complementary wedges, the ideal conductor boundary conditions will count
the azimuthally constant TE mode twice whereas with more realistic boundary
conditions such as considered here, such a mode must be common to the
\emph{both sectors}, $0\leq\theta<2\pi$. In these two respects the perfectly
conducting wedge differs from the diaphanous one, and put together these
redefinitions provided by the diaphanous wedge exactly remove the divergent
extra energy term found in \cite{brevik09} and previously in
\cite{nesterenko03}.

We show numerically that except for the singular term, the energy of a
perfectly conducting wedge closed by a magnetodielectric cylinder whose
reflectivity tends to unity is regained in the limit where we let the wedge
boundaries become perfectly reflecting.

\section{Boundary conditions and dispersion relations}

\begin{figure}[tb]
  \begin{center}
    \includegraphics[width=1.7in]{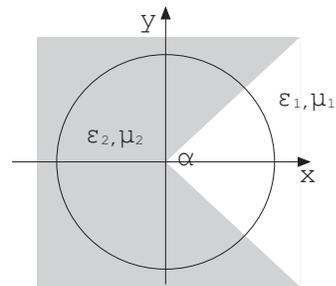}
    \caption{The wedge geometry considered.}
    \label{fig_wedge}
  \end{center}
\end{figure}

We begin by considering in general the form of an expression of the energy
of a diaphanous wedge inside and outside a cylindrical shell such as depicted
in Fig.~\ref{fig_wedge}. We assume the cylindrical shell to be
perfectly reflecting.
Let the interior sector $-\alpha/2<\theta<\alpha/2$ have permittivity and
permeability $\epsilon_1$ and $\mu_1$ relative to vacuum, and the
corresponding values for the exterior sector $\pi\ge|\theta|>\alpha/2$
be $\epsilon_2$ and $\mu_2$ so that $\epsilon_1(\omega)\mu_1(\omega)=
\epsilon_2(\omega)\mu_2(\omega)\equiv n^2(\omega)$. The cusp of the wedge is
chosen to lie along the $z$ axis, which is also the center of the cylindrical
shell, and the interfaces are found at $\theta=\pm\alpha/2$ and at $\rho=a$
($\rho$ is the distance to the $z$ axis).

We will calculate the Casimir energy by `summing' over the eigenmodes of the
geometry using the so-called argument principle, now a standard method in the
Casimir literature. The eigenmodes of a given geometry are given by the
solutions of the homogeneous Helmholtz equation
\be\label{helmholz}
  (\nabla^2-n^2 \partial_t^2)u(\br,t) =0,
\ee
which also satisfy the system's boundary conditions. Here $u$ symbolises a
chosen field component of either the electric or magnetic field. We will
choose $E_z$ and $H_z$ as the two independent field components from which the
rest of the components can be derived by means of Maxwell's equations.

The translational symmetry with respect to $z$ and time makes it natural to
introduce the Fourier transform
\[
  E_z(\br,t) = \int_{-\infty}^\infty \frac{d\omega}{2\pi}
e^{-i\omega t}\int_{-\infty}^\infty \frac{dk_z}{2\pi}e^{ik_zz}
E_z(\boldsymbol\rho;\omega,k_z),
\]
where $\boldsymbol\rho=(\rho,\theta)$ and $\rho=\sqrt{x^2+y^2}$. The
Helmholtz equation now simplifies to the scalar Bessel equation
\be
  (\nabla^2_\perp + \kp^2)E_z\Arkw = 0,\label{HhHom}
\ee
where
\be
  \nabla_\perp^2 = \partial_\rho^2 + \frac1{\rho}\partial_\rho
+ \frac1{\rho^2}\partial_\theta^2
\ee
and $\kp^2 = \epsilon\mu\omega^2 - k_z^2$.

We will define the quantity $\kappa$ as
\be
  \kappa = \sqrt{k_z^2-\epsilon\mu\omega^2} = -i\kp,
\ee
where the root of $\kappa$ is to be taken in the fourth complex quadrant.
When in the end we take frequencies to lie on the positive imaginary axis,
$\kappa$ becomes real and positive, something we bear in mind in the
subsequent calculations.

A general solution to Eq.~(\ref{HhHom}) is of the form
\be\label{gen_sol}
  E_z = [A_\nu\Hn(\kp \rho) + B_\nu \Jn(\kp \rho)]
(ae^{i\nu\theta}+be^{-i\nu\theta}),
\ee
where $A_\nu,B_\nu,a,b$ and $\nu$ are arbitrary. If, as in our case, $\nu$
is allowed to take non-integer values, we must restrict it to $\nu\geq0$
because except at integers $J_\nu(z)$ and $J_{-\nu}(z)$ are linearly
independent.

Solutions of the EM field in a wedge geometry are expressed as a sum over
cylindrical partial waves whose kernel are Bessel and Hankel functions of
argument $\kp \rho$. Thus it is clear that the boundary conditions on the
wedge surfaces can only be solved for each partial wave if the speed of
light is the same in both sectors since $\kp$ would otherwise take
different values in the two media for given $k_z$ and $\omega$ and the
kernel functions would be linearly independent functions of these. The
diaphanous condition is thus prerequisite for the explicit solution of
boundary conditions below. Without this condition the problem at hand is
not analytically solvable with the methods used herein.  We expect that
even if we could solve the nondiaphanous problem we would encounter 
divergences that might or might not be curable by the inclusion of 
dispersion.

The presence of the wedge primarily has the r\^{o}le of dictating which
values of $\nu$ are allowed. If one were to consider a cylinder (periodic
boundary conditions), only integer values of $\nu$ (both positive and
negative) would be acceptable and expressing the solution as a sum over these
integer values would be appropriate. Were one instead to let the wedge be
perfectly reflecting (Dirichlet and Neumann boundaries at $\pm\alpha/2$
where $\alpha$ is arbitrary) $\nu$ would be forced to take values that are
non-negative integer multiples of $\pi/\alpha$. The diaphanous
magnetodielectric boundaries present here also restrict $\nu$ to discrete
values for given $\epsilon$'s and $\mu$'s, but explicitly determining these
values is no longer immediate because modes existing in the exterior and
interior sectors now couple to each other. For a given frequency we therefore
make use of an appropriate dispersion function representing these boundaries
in order to sum over the appropriate values of $\nu$ by means of the argument
principle, whereupon we may sum over the eigenfrequencies of the modes inside
and outside the cylindrical shell to obtain the energy.

The boundary condition dispersion relation pertaining to the circular
boundary is known (e.g.\ \cite{brevik09}, Eq.\ (4.12), with $\xi^2=1$),
\be\label{circle}
  g_\nu(k_z, \omega) \equiv 1- x^2 \lambda_\nu^2(x) = 0,
\ee
where $x = a\kappa$,
\be
  \lambda_\nu(x) = \frac{d}{dx}[I_\nu(x)K_\nu(x)],
\ee
$I_\nu, \Kn$ are the modified Bessel functions of the first and second kind
of order $\nu$. We can simply use this equation to sum modes satisfying the
boundary condition on both sides because the wedge boundaries at
$\pm\alpha/2$ impose the same discretisation of $\nu$ inside and outside the
cylindrical shell (were we to have e.g.\ a third medium in the sector
$|\theta|<\alpha/2, \rho>a$ different from medium $1$, this would no longer
be the case as we will see: the field solutions would take different values
of $\nu$ inside and outside the cylindrical boundary and the boundary
conditions at the cylinder could no longer be solved for each eigenvalue of
$\nu$). We now turn to a derivation of the dispersion relation pertaining to
the interfaces at $\theta=\pm\alpha/2$.

In the following we shall use the term TE to denote electromagnetic modes
whose $\mathbf{E}$ field has no component in the $z$ direction and TM
denotes those modes whose $\mathbf{H}$ field has no $z$ component. This is
not `transverse electric' and `transverse magnetic' with respect to the wedge
boundaries at $\theta=\pm\alpha/2$, but this does not matter since we will
find that the eigenequation of these boundaries is the same for all field
components by virtue of the diaphanous condition.

\subsection{Kontorovich-Lebedev approach}

We will first employ the technique of the Kontorovich-Lebedev (K-L)
transformation \cite{kontorovich38} and its inverse transform which may be
written as:
\begin{subequations}
\begin{align}
  E_z(\boldsymbol\rho) =& i\int_0^{i\infty} d\nu\nu e^{\frac{i\nu\pi}{2}}
\sin(\pi\nu)\Kn(\kappa \rho)\gE_z(\theta; \nu),\label{WEKL}\\
  \gE_z(\theta; \nu) =& \frac{2}{\pi^2}\int_0^\infty \frac{d \rho}
{\rho}e^{-\frac{i\nu\pi}{2}}\Kn(\kappa \rho)E_z(\boldsymbol\rho), \label{WeKL}
\end{align}
\end{subequations}
(dependence on $k_z$ and $\omega$ is implicit). While less extensively
covered in the literature than most other integral transforms, some tables of
K-L transforms exist \cite{BookDitkin65,BookOberhettinger72}. Numerical
methods for evaluating such transforms were recently developed by Gautschi
\cite{gautschi06}. We will ignore the presence of the cylindrical shell in
this section and only study how the presence of the walls of the wedge
discretize the spectrum of allowed values of the Bessel function order
$\nu$.

With this, (\ref{HhHom}), after multiplying with $\rho^2$, transforms to
\be\label{HelmholzNu}
  (\partial_\theta^2 + \nu^2)\gE_z\Antkw = 0.
\ee
Equation (\ref{HelmholzNu}) is now in a form fully analogous to that
encountered in a planar geometry (e.g.~\cite{schwinger78, ellingsen07,
tomas95}). We follow now roughly the scheme of \cite{tomas95} and determine
the dispersion relation [condition for eigensolutions of Eq.~(\ref{HhHom})] by
means of summation over multiple reflection paths. By noting that the
solutions to Eq.~(\ref{HelmholzNu}) have the form of propagating plane waves
travelling clockwise or anticlockwise along the now formally straight
$\theta$ axis ($\nu$ playing the role of a reciprocal azimuthal angle) the
analogy to a plane parallel system is obvious.

We write the solution of Eq.~(\ref{HelmholzNu}) in the interior sector,
$|\theta|<\alpha/2$, in the form
\be\label{ezKLh}
  \gE_z = e^+ e^{i\nu\theta} + e^-e^{-i\nu\theta},
\ee
where $e^\pm$ are undetermined integration coefficients, field amplitudes at
$\theta=0$ to be determined from boundary conditions at $\theta=\pm\alpha/2$.

Likewise the solutions in the exterior sector (the `complementary wedge')
$\pi\ge|\theta|>\alpha/2$ may be written
\be
  \tgE_z = \te^+ e^{i\nu(\theta-\pi)} + \te^- e^{-i\nu(\theta-\pi)},
\ee
where the undetermined amplitudes $\te^\pm$ are `measured' at $\theta=\pi$.
The choice to measure the amplitudes in sectors 1 and 2 at $\theta=0$ and
$\pi$ respectively is arbitrary, but makes for maximally symmetric boundary
equations.

The homogeneous Helmholtz equation thus solves the scattered part of $\gE_z$
given some source field $\gE^0_z$. Let us assume there is a source field in
form of an infinitely thin phased line source parallel to the $z$ axis at
some position  $\theta_0$ in the interior sector. The direct field (which
only propagates away from the source) may be written in the form
\be
  \gE_z^0 = \Theta(\theta-\theta_0)e_0^+e^{i\nu\theta}
+\Theta(\theta_0-\theta)e_0^-e^{-i\nu\theta},
\ee
where $\Theta(x)$ is the unit step function and the field amplitudes are
``measured'' at $\theta=0$. We do not need to know the constants $e_0^\pm$
explicitly and take these to be known constants. The multiple reflection
problem (or equivalently, boundary condition problem) is now a system of four
equations for the four amplitudes $e^\pm, \te^\pm$ as functions of $e_0^\pm$.

We define the reflection coefficients at the boundaries $\theta=\pm\alpha/2$
as the ratio of reflected vs.\ incoming field amplitude,
$r= \gE_{z,\text{refl}}/\gE_{z,\text{in}}$, as seen by a wave coming from and
reflected back into sector 1 (a wave going the opposite way experiences a
coefficient $-r$). With the assumption $\epsilon_1\mu_1=\epsilon_2\mu_2$ the
reflection coefficients of the $s$ and $p$ modes differ only by a sign:
\be
  r_p = \frac{\epsilon_2-\epsilon_1}{\epsilon_2+\epsilon_1} = -r_s
=-\frac{\mu_2-\mu_1}{\mu_2+\mu_1}. \label{diaphref}
\ee
We will simply use $r$ in the following, representing either of the modes. We
also define the transmission coefficient, the ratio of the transmitted to the
incoming amplitude, going from sector $i$ to sector $j$, $t_{ij}$ where
$i,j=1,2$ denotes the sectors in figure \ref{fig_wedge},
\be
  t_{ij}\equiv t_{ij,s} = \frac{\mu_j}{\mu_i}\frac{2\epsilon_j}
{\epsilon_j+\epsilon_i} = t_{ij,p} = \frac{2\mu_j}{\mu_j+\mu_i}.
\ee
Since these coefficients are invariant under K-L transformation, they are the
sought-after single interface reflection and transmission coefficients also
in the K-L regime. Note that with the diaphanous condition, reflection
coefficients are independent of $\nu$, something which would not be true in
general. Were $r$ to depend on $\nu$ this would give rise to corrections to
the energy expression derived in section \ref{sec_non-disp}.  (See also
the Appendix, where such $\nu$ dependence does occur.)

\begin{figure}[tb]
  \begin{center}
  \includegraphics[width=3in]{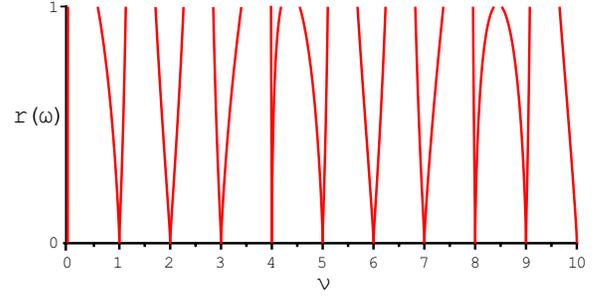} 
  \caption{The solutions of the dispersion relation (\ref{DEfix}) as a
function of $r$ and $\nu$ for $\alpha=0.75$. The eigenvalues of $\nu$ for a
given $r$ are marked; the energy is calculated by summing over these
values and then integrating over all $\omega$.}
  \label{fig_disp}
  \end{center}
\end{figure}

We formulate the electromagnetic boundary conditions in terms of reflection
and transmission. In the K-L domain the system looks and behaves analogously
to the planar system (see \cite{tomas95} for details on this case), but with
one important difference, namely that a $\theta$ directed partial wave which
is transmitted at a wedge boundary does not disappear from the system but is
partly transmitted back into sector 1 again circularly. Thus we obtain four
equations for the four amplitudes $e^+, e^-,\te^+$ and $\te^-$, coupling to
each other through paths reflected or transmitted at one interface:
\begin{subequations}
  \begin{align}
e^+ =& r e_0^- e^{i\nu\alpha} + r e^- e^{i\nu\alpha} + t_{21}\te^+ e^{i\nu\pi},
\\
e^- =& r e_0^+ e^{i\nu\alpha} + r e^+ e^{i\nu\alpha} + t_{21}\te^- e^{i\nu\pi},
\\
\te^+ =& t_{12}e_0^+ e^{i\nu\pi}+t_{12}e^+ e^{i\nu\pi} - r\te^-
e^{i\nu(2\pi-\alpha)},\\
\te^- =& t_{12}e_0^- e^{i\nu\pi}+t_{12}e^- e^{i\nu\pi}
- r\te^+ e^{i\nu(2\pi-\alpha)}.
  \end{align}
\end{subequations}

Eigenvalues of $\nu$ for the wedge correspond to solutions of these boundary
conditions, which exist when the secular equation of the set of linear
equations for $e^\pm$ and $\te^\pm$ is fulfilled. The characteristic matrix is
\be
  \matD=\left(\begin{array}{cccc}
    1&-re^{i\nu\alpha}&-t_{21}e^{i\nu\pi} &0 \\
    -re^{i\nu\alpha}&1&0&-t_{21}e^{i\nu\pi} \\
    -t_{12}e^{i\nu\pi}&0&1&re^{i\nu(2\pi-\alpha)}\\
    0&-t_{12}e^{i\nu\pi}&re^{i\nu(2\pi-\alpha)}&1 \end{array}\right)
\label{matrix}
\ee
and the dispersion relation sought after is
\be\label{dispersion}
   D(\nu,\omega)\equiv \det\matD=0.
\ee
The matrix form (\ref{matrix}) is rather instructive. Note how $\matD$ is a
block matrix of the form
\[
  \matD = \left(\begin{array}{cc}\mathbf{D}_{1} &\mathbf{G}_{21} \\
\mathbf{G}_{12} &\mathbf{D}_{2}\end{array}\right),
\]
where $\mathbf{D}_i$ describes multiple scattering within sector $i$ and
$\mathbf{G}_{ij}$ describes coupling between the sectors by transmission from
sector $i$ to $j$. Since the $\mathbf{G}$ matrices commute with the
$\mathbf{D}$ matrices, $\det\matD$ can be written
\be\label{detIdent}
\det\matD = \det (\mathbf{D}_{1}\mathbf{D}_{2} - \mathbf{G}_{21}
\mathbf{G}_{12}).
\ee
We may use the energy conservation relation
\be\label{rrtt}
   t_{12}t_{21} + r^2 = 1,
\ee
together with (\ref{detIdent}) to find the simple expression
\begin{align}\label{DEfix}
  D(\nu,\omega) =& (1-e^{2\pi i\nu})^2- r^2(e^{i\nu(2\pi-\alpha)}-
e^{i\nu\alpha})^2\notag \\
  =& -4e^{2\pi i\nu}[\sin^2(\nu\pi)- r^2\sin^2(\nu(\pi-\alpha))].
\end{align}
It is noteworthy that this dispersion relation only has an implicit
dependence on $\omega$ through the quantity $r^2(\omega)$.  As an example we
plot the
solutions to Eq.~(\ref{dispersion}) as a function of $\nu$ and $r$ in
Fig.~\ref{fig_disp} for $\alpha=0.75$ radians.

Note at this point that whenever $r$ is real, all zeros of $D(\nu,\omega)$ in
(\ref{DEfix}) are real. In the following we shall think of $r$ as well as the
eigenvalues of $\nu$ as real quantities. For real frequencies $\omega$
reflection coefficients will in general be complex, while after a standard rotation of frequencies onto the imaginary
frequency axis these coefficients are always real as dictated by
causality. Although zeros are complex the argument principle may still be used;
the discussion of connected subtleties may be found in e.g.\ \cite{deraad81,brevik01b,nesterenko06}.

It is easy to see that this dispersion relation generalises that for a
cylinder (of infinite radius) and a perfectly conducting wedge. In the latter
limit, $r=1$, the determinant $\det\matD$ has zeros where $\nu=m \pi/\alpha$
and at $\nu = m \pi/(2\pi-\alpha)$ where $m$ is an integer. This becomes
obvious when noting that
\be\label{perf_ref}
D(\nu,\omega) \buildrel{r\to 1}\over{\longrightarrow}
-4e^{2\pi i\nu}\sin\nu\alpha\cdot\sin\nu(2\pi-\alpha).
\ee
This reproduces, in other words, the case where the wedge is made up of thin
perfectly conducting sheets. For the perfectly conducting wedge it is
customary to restrict $\nu$ to values that are integer multiples of
$\pi/\alpha$ from the beginning.

Likewise when the two materials become equal,
\be\label{cylinder}
  D(\nu,\omega) \buildrel{r=0}\over{\longrightarrow}
-4e^{2\pi i\nu}\sin^2\nu\pi \equiv D_0(\nu),
\ee
which has double zeros where $\nu=m$, a positive integer, corresponding to a
clockwise and an anticlockwise mode or, if the reader prefers, the sum over
$\nu=+m$ and $-m$. This is just the cylinder case \cite{deraad81,brevik94,
gosdzinsky98,milton99, lambiase99, caveroPelaez05, romeo05,brevik07}. We see
from Fig.~\ref{fig_disp} that except for $\nu=0$ which remains degenerate,
the double zeroes split into two separate simple zeros for finite $r$. For
special opening angles which are rational multiples of $\pi$ there will be
other zeroes which remain degenerate as well.

One sees directly that were we to solve Eq.~(\ref{HhHom}) for $H_z$ instead of
$E_z$ the dispersion relation would be identical to Eq.~(\ref{dispersion})
since
the only difference would be the sign of the reflection coefficient (we would
employ $r_s$ rather than $r_p$), which only enters squared. One should note
that the distinction between $r_s$ and $r_p$ here does not correspond to the
distinction between TE and TM modes of the entire cavity, but this is of no
consequence in the following because the dispersion relation
(\ref{dispersion}) is the same for all field components!

\subsection{Derivation by standard expansion}\label{sec_standard}

We will now sketch how the result (\ref{DEfix}) may be derived by a more
standard method similar to that made use of in \cite{brevik09}. The solutions
of Eq.~(\ref{gen_sol}) that correspond to outgoing waves at
$\rho\to \infty$ may
be expanded following the scheme of \cite{brevik09} in an obvious
generalisation of those found in \cite{BookStratton41}. Due to criteria of
outgoing-wave boundary conditions at $\rho\to\infty$ and non-singularity at
the origin the solution must consist purely of Hankel function
$H^{(1)}_\nu(\kp\rho)$ far from the origin and only of terms containing
$J_{\nu}(\kp\rho)$ near $\rho=0$. Following the scheme of \cite{brevik09} we
choose $H_\nu^{(1)}(\kp\rho)$ for $\rho\geq a$ and $J_\nu(\kp\rho)$ for
$\rho\leq a$ both in the interior sector $-\alpha/2<\theta<\alpha/2$ and
outside, and couple the solutions across these straight boundaries. It will
not matter which Bessel function we choose for the present purposes: the
resulting solution expansions are identical but for the replacement of one
Bessel function with another.

In a straightforward generalisation of the expansion used in \cite{brevik09}
we write down the following general solutions in sector 1 of
Fig.~\ref{fig_wedge} for $\rho>a$:
\begin{widetext}
  \begin{subequations}
    \begin{align}
      E_{r,1}=&\int_0^\infty d\nu \left\{\left[ \frac{i\kz}{\kp}\Hnp\oa_1
-\frac{\nu\mu_1\omega}{\kp^2\rho}\Hn\ub_1 \right]\cos\nu\theta
+ i\left[ \frac{i\kz}{\kp}\Hnp\ua_1
-\frac{\nu\mu_1\omega}{\kp^2\rho}\Hn\ob_1\right]\sin\nu\theta\right\}
e^{i\pi\nu/2},\\
      E_{\theta,1}=&-\int_0^\infty d\nu \left\{\left[
\frac{\nu\kz}{\kp^2\rho}\Hn\ua_1+\frac{i\mu_1\omega}{\kp}\Hnp\ob_1 \right]
\cos\nu\theta + i\left[\frac{\nu\kz}{\kp^2\rho}\Hn\oa_1
+\frac{i\mu_1\omega}{\kp}\Hnp\ub_1\right]\sin\nu\theta\right\}e^{i\pi\nu/2},\\
  E_{z,1}=&\int_0^\infty d\nu \Hn\left[\oa_1 \cos\nu\theta
+ i\ua_1 \sin\nu\theta\right]e^{i\pi\nu/2},\\
      H_{r,1}=&\int_0^\infty d\nu \left\{\left[
\frac{\nu\omega\epsilon_1}{\kp^2\rho}\Hn\ua_1
+\frac{i\kz}{\kp}\Hnp\ob_1 \right]\cos\nu\theta
+ i\left[ \frac{\nu\omega\epsilon_1}{\kp^2\rho}\Hn\oa_1+\frac{i\kz}{\kp}
\Hnp\ub_1\right]\sin\nu\theta\right\}e^{i\pi\nu/2},\\
      H_{\theta,1}=&\int_0^\infty d\nu \left\{\left[
\frac{i\omega\epsilon_1}{\kp}\Hnp\oa_1-\frac{\nu\kz}{\kp^2\rho}\Hn\ub_1
\right]\cos\nu\theta + i\left[ \frac{i\omega\epsilon_1}{\kp}\Hnp\ua_1
-\frac{\nu\kz}{\kp^2\rho}\Hn\ob_1\right]\sin\nu\theta\right\}e^{i\pi\nu/2},\\
      H_{z,1}=&\int_0^\infty d\nu \Hn\left[\ob_1 \cos\nu\theta
+ i\ub_1 \sin\nu\theta\right]e^{i\pi\nu/2},
    \end{align}
   \end{subequations}
\end{widetext}
where we have omitted the arguments of $\Hna$ and its derivative, and of
$\oa_1(\nu), \ob_1(\nu)$, etc., the latter being undetermined coefficient
functions of $\nu$.

We write the solution in sector 2 in exactly the same form but with the
simple replacements $\theta\to \theta-\pi$ and  $\epsilon_1\to \epsilon_2,
\mu_1\to\mu_2$ and the same for the coefficient functions. With the
isorefractive assumption $\kp$ is the same in both media for given $\omega$
and $\kz$, so the boundary conditions at the interfaces can be solved under
the integral signs. In general there are 8 unknown functions and eight
equations, yet one finds that the $s$ and $p$ modes decouple into linear
equation sets of $4\times4$ on the form
\be
  \mattD\cdot \mathbf{a}=0
\ee
where  $\mattD$ equals
\[
  \left(\begin{array}{cccc}
    \cos\frac{\nu\alpha}{2}& 0 & -\cos\nu(\frac{\alpha}2-\pi) &0 \\
    0 & \sin\frac{\nu\alpha}{2} &0& - \sin\nu(\frac{\alpha}2-\pi) \\
    -\epsilon_1\sin\frac{\nu\alpha}{2} &0& \epsilon_2
\sin\nu(\frac{\alpha}2-\pi) &0\\
    0&-\epsilon_1\cos\frac{\nu\alpha}{2} &0& \epsilon_2
\cos\nu(\frac{\alpha}2-\pi)
  \end{array}\right)
\]
and $\mathbf{a}$ is a vector, either $(\oa_1, \ua_1, \oa_2, \ua_2)$ or
$(\ob_1, \ub_1, \ob_2, \ub_2)$.

As before the eigenmodes of the system solve the equation $\det\mattD=0$.
With some manipulation we find that the determinant can be written simply as
\begin{align}
  \det\mattD =& \frac1{4}(\epsilon_2-\epsilon_1)^2\sin^2\nu(\pi-\alpha)
\notag \\
  &- \frac1{4}(\epsilon_2+\epsilon_1)^2\sin^2\nu\pi.\label{sinDisp}
\end{align}
Under the assumption that $\epsilon_2+\epsilon_1\neq 0$ the equation
$\det\mattD=0$ is equivalent to Eq.~(\ref{dispersion}) with Eq.~(\ref{DEfix}).

\section{The Casimir energy}

In order to find the Casimir energy we shall employ the argument principle,
introduced to the field of Casimir energy by van Kampen et al.\
\cite{vankampen68} who rederived Lifshitz's result in a simple way. For a
very readable review of the technique, see \cite{BookParsegian06}.

A similar system to that shown in Fig.~\ref{fig_wedge} was considered in
\cite{brevik09} where the plane sides of the wedge were instead made up of
perfectly reflecting interfaces and the circular boundary was diaphanous. We will start from the result of \cite{brevik09} and generalise this step by step to approach the desired energy expression for the current situation.
Excepting the formally singular energy term associated with the sharp corners
where the arc meets the wedge walls found in that paper (we shall regard
this term separately below), the Casimir energy per unit length of that system
in the limit of perfectly reflecting circular arc was (Eq.\ (2.10) or
(4.11) of \cite{brevik09})
\be \label{E_perfWedge}
  \tmE_\text{id} = \frac{1}{2\pi i}\int_{-\infty}^\infty
\frac{dk_z}{2\pi} \msum \oint_\Lambda d\omega\frac{\omega}{2}
\frac{d}{d\omega}\ln g_{mp}(k_z,\omega),
\ee
with $g_\nu(k_z,\omega)$ given in Eq.~(\ref{circle}) and we define the
shorthand
\be
  p = \frac{\pi}{\alpha}.
\ee
The prime on the summation mark means the $m=0$ term is taken with half
weight. The integration contour $\Lambda$ is chosen to follow the imaginary
axis and is closed to the right by a large semicircle thus encircling the
positive real axis. The roots of (\ref{circle}) are in general
complex; the applicability of the argument principle for such situations was
discussed in \cite{deraad81,brevik01b, nesterenko06,nesterenko08}. The energy
$\tilde{\mathcal E}$ has been normalised so as to be zero when the circular
arc is moved to infinity.

Each frequency satisfying $g_{mp}(k_z,\omega)=0$ gives a pole which adds the
zero temperature energy $\frac{\omega}{2}$ of that mode through Cauchy's
integral theorem. In the end there are sums over the eigenvalues of $\nu$,
$mp$, the eigenvalues found when the sides of the wedge are assumed to be
perfectly conducting. Employing such an assumption from the start completely
decouples the interior sector $|\theta|<\alpha/2$ from the exterior. If we
were to interpret the perfectly reflecting wedge as the \emph{limit} of an
isorefractive wedge such as that described by the dispersion relation
(\ref{DEfix}) as $|r|\to 1$, however (for example by letting $\epsilon_2\to
\infty$ and $\mu_2\to 0$ so that their product is constant), the interior and
exterior sectors remain coupled and we obtain an additional $m$-sum, namely
that over $\nu=mp'$ of the complementary wedge, where
\be\label{pprime}
  p' = \frac{\pi}{2\pi-\alpha} = \frac{p}{2p-1}.
\ee

To obtain direct correspondence we therefore modify Eq.~(\ref{E_perfWedge}) by
also including the energy of the modes of the complementary wedge, fulfilling
$\nu=mp'$. Since we will soon generalise this result to the case where the
wedge is diaphanous, it is reasonable to subtract the energy
corresponding to absence of the boundaries at $\pm\alpha/2$, by subtracting
off the energy obtained were $\nu$ to fulfill periodic boundary conditions
(i.e.\ a circular cylinder). The result is
\be\label{E_perfMod}
  \tmE_\text{id} \to \frac{1}{4\pi i}\int_{-\infty}^\infty \frac{dk_z}{2\pi}
\msum \oint_\Lambda d\omega\omega \frac{d}{d\omega}\ln \frac{g_{mp}\cdot
g_{mp'}}{g_m^2}.
\ee
The periodic function $g_m(k_z,\omega)$ is squared since both positive and
negative integer orders contribute equally in the periodic case, and the
symmetry under $m\to-m$ makes for a factor of 2 except for $m=0$. The latter
exception is automatically accounted for by the prime on the sum.

Note that employing $g_\nu(k_z,\omega)$ with the argument principle
automatically takes care of the sum over the two polarisations since, by
virtue of the diaphanous condition, $g_\nu$ is a product of boundary
conditions for TE and TM modes (see e.g.\ Appendix B of \cite{brevik09}).

Let us now perform the generalisation of Eq.~(\ref{E_perfMod}) to the present
case. The sum over $\nu=mp$ and $mp'$ may be generalised to a sum over the
solutions of Eq.~(\ref{dispersion}) using the argument principle once more to
count the zeros of Eq.~(\ref{DEfix}), and the subtraction of the periodic modes
in the absence of the boundary is performed by subtracting the solutions of
$D_0(\nu)=0$ with $D_0$ from Eq.~(\ref{cylinder})
(note that the zeros of $D_0$
are double, automatically giving the factor 2 manually introduced in
Eq.~(\ref{E_perfMod}) by taking the square of $g_m$). We obtain
\begin{align}
    \tmE =& \frac{1}{2(2\pi i)^2}\int_{-\infty}^\infty \frac{dk_z}{2\pi}
\oint_\Lambda d\omega\omega \notag \\
    &\times\oint_\Lambda d\nu  \left[\frac{d}{d\omega}\ln g_\nu(k_z,\omega)
\right]\frac{d}{d\nu}\ln \frac{D(\nu,\omega)}{D_0(\nu)}.
\end{align}
The contour of the $\nu$ integral is the same as that for the $\omega$
integral.

Neither of the contour integrals obtain contributions from the semicircular
contour arcs so we are left with integrals over imaginary order and frequency.
Performing substitutions $\omega=i\zeta$ and $\nu = i\eta$ we obtain
\begin{align}
    \tmE=& \frac{1}{16\pi^3 i}\int_{-\infty}^\infty dk_z \int_{-\infty}^\infty
d\zeta\zeta\notag \\
    &\times\int_{-\infty}^\infty d\eta  \left[\frac{d}{d\zeta}
\ln g_{i\eta}(k_z,i\zeta)\right]\frac{d}{d\eta}\ln \frac{ D(i\eta,i\zeta)}
{D_0(i\eta)}.\label{fullEnergy}
\end{align}
This is the general form of the Casimir energy of the system presented herein.

To be very explicit about the regularisations performed: Eq.~(\ref{fullEnergy})
is the energy of the geometry of Fig.~\ref{fig_wedge}, minus the energy
when the cylinder is pushed to infinity (the double wedge alone), minus the
renormalised energy of the cylinder relative to uniform space,
\be
  \tmE = (\mE_\olessthan - \mE_<) - \tmE_\circ,
\ee
where $\tmE_\circ$ is the $\zeta$-renormalised energy of a cylindrical shell
(relative to uniform space) considered in \cite{milton99}, and $\olessthan,
<$ symbolise the double wedge with and without the cylindrical shell. It is
thus clear that the energy should vanish when either the cylindrical boundary
tends to infinity ($\mE_\olessthan \to \mE_<$ and $\tmE_\circ\to 0$) or when
the wedge becomes completely transparent ($\mE_\olessthan - \mE_< \to
\tmE_\circ$).

The corresponding free energy at finite temperature $T$ is found by simply
substituting the integral over $\zeta$ in Eq.~(\ref{fullEnergy}) with the well
known Matsubara sum over the frequencies $\zeta_k = 2\pi k T$ where
$k\in\mathbb{Z}$;
\be
  \int_{-\infty}^\infty d\zeta f(i\zeta)\to 2\pi T\sum_{k=-\infty}^\infty
f(i\zeta_k).
\ee
We will not consider finite temperature numerically in the present effort.

\subsection{Non-dispersive approximation}\label{sec_non-disp}

In order to proceed to producing numerical results we make the simplifying
 assumption that $r$ be approximately constant with respect to $\zeta$ over
the important range of $\zeta$ values; $\frac{dr}{d\zeta}\approx 0$.
This a version of the constant reflection coefficient model which was
previously found to be useful for the planar geometry \cite{ellingsen09}.
While it is true that for any \emph{real} material, reflectivity must tend
to zero at infinite frequency, the non-dispersive approximation is a useful one
and allows a simpler expression to be derived.
We will find below that the resulting Casimir energy expression is finite even
when $r=1$ for all frequencies, except when $\alpha=0$ or $2\pi$. There is
therefore no need to assume high-frequency transparency for the sake of
finiteness in this case.

With this assumption we can easily perform a partial integration in $\zeta$. 
We note that, when
$r$ is independent of $\nu$ as in the diaphanous case (see the Appendix for a
situation where this is not so),
\begin{align}
 \frac{d}{d\eta} \left[\ln\frac{D}{D_0}\right]
=& \frac{\alpha \sinh\eta(2\pi-\alpha)-(2\pi-\alpha)\sinh\eta\alpha}
{\sinh^2\eta\pi-r^2\sinh^2\eta(\pi-\alpha) }\notag \\
  &\times\frac{r^2 \sinh\eta(\pi-\alpha)}{\sinh\eta\pi},
\end{align}
which is now approximated as independent of $\zeta$ and $k_z$. It is then
opportune to perform a change of integration into the polar coordinates
\be
  X = n\zeta=\kappa\cos\theta;~~~Y = k_z = \kappa\sin\theta,\label{pcoord}
\ee
so that $X^2+Y^2=\kappa^2$ and
\be
  \int_{-\infty}^\infty dk_z \int_{-\infty}^\infty d\zeta f(a\kappa)
= \frac{2\pi}{na^2}\int_0^\infty dxx f(x),
\ee
where $x=a\kappa$ as before. We obtain after integrating by parts
\begin{widetext}
  \be \label{BigEnergy}
    \tmE=\frac{i}{8\pi^2 na^2 }\int_{-\infty}^\infty d\eta
\frac{r^2 \sinh\eta(\pi-\alpha)[\alpha \sinh\eta(2\pi-\alpha)
-(2\pi-\alpha)\sinh\eta\alpha]}{\sinh\eta\pi[\sinh^2\eta\pi
-r^2\sinh^2\eta(\pi-\alpha)]}\int_0^\infty dx x\ln[1-x^2\lambda_{i\eta}^2(x)].
  \ee
\end{widetext}
Despite appearances this expression is in fact real. This is because the
dispersion function in the first integral is an odd function of $\eta$
while the real and imaginary parts of the logarithm are even and odd
respectively (provided the appropriate branch of the logarithm is taken),
hence the imaginary part of $\tmE$ vanishes under symmetrical integration.
It is straightforward to write down the correction terms containing
$\frac{dr}{d\zeta}$ or $\frac{dr}{d\nu}$ should the reader wish to do so.
Such is necessary were
one to study the role of dispersion on the energy; we shall not consider this
herein---but see the Appendix for $dr/d\nu\ne0$.

The energy expression (\ref{BigEnergy}) has the reasonable properties of
being zero at $\alpha=\pi$ and symmetrical under the substitution
$\alpha\leftrightarrow 2\pi-\alpha$. We will study Eq.~(\ref{BigEnergy})
numerically in section \ref{sec_numerics}. We argue in the next subsection
that Eq.~(\ref{fullEnergy}) is the full Casimir energy of this system (after
subtracting that of the cylinder alone). Thus the zero energy at
$\alpha=\pi$ demonstrates a particular generalisation of the theorem of
Ambj\o rn and Wolfram (\cite{ambjorn83}, stated in Eq.~(2.49) of
\cite{BookMilton01}): the energy of a semi-circular compact diaphanous
cylinder is half that of a full cylinder (there is an equal contribution from
the exterior `half-cylinder' so the difference is zero).

For large $\eta$ the term proportional to $\alpha$ in the big fraction in
Eq.~(\ref{BigEnergy}) behaves for $\pi-\alpha>0$ as
\be
  \frac{d}{d\eta}\ln\frac{D}{D_0}\sim \frac{2\alpha r^2}{e^{2\alpha\eta}-r^2},
\ee
with a similar behavior for the term proportional to $2\pi-\alpha$,
and so is exponentially convergent. With perfect reflectivity
Eq.~(\ref{BigEnergy}) is finite except when  $\alpha$ equals $0$ or $2\pi$
when $|r|=1$.

\subsection{No additional corner term}\label{sec_addterm}

In the geometry considered in \cite{brevik09}, which differed from the
present one primarily by the assumption that the wedge be perfectly conducting,
the Casimir energy was found to possess a divergent term which could be
associated with the corners where the arc meets the wedge sides. When the arc
was instead made diaphanous it was shown that this term could be rendered
finite by virtue of high frequency transparency as displayed by any real
material boundary.

The energy (\ref{fullEnergy}) is the direct generalisation of the finite
part of the energy of the system considered in \cite{brevik09}. We will argue
that when the wedge is also diaphanous, this is indeed the full energy of the
system, regularised by the subtraction of the energy of the cylinder alone
(which in turn is regularised by subtracting the energy of uniform space).

Let us recapitulate how the divergent term in \cite{brevik09} came about. The
zeta function regularised energy expression (Eq.~(4.13) of
\cite{brevik09}) adds the $m=0$ modes of both polarisations with half weight.
There should be no $m=0$ TM mode, however, because the perfectly conducting
wedge forces any azimuthally constant electric field to have zero amplitude
everywhere, thus the half-weight zero TM mode should be subtracted. Moreover, 
since for
arbitrary opening angles only positive values of $m$ are allowed, the zero TE
mode should be counted with full rather than half weight, and thus the
correction term equals one half the $m=0$ energy of the TE mode minus one half 
that of the TM mode.

In contrast we are here not considering perfectly conducting wedge boundaries
so the
TM $m=0$ mode should be included. The question becomes whether the $\nu=0$ TE
and TM modes have been counted with only half the weight they should. In a
system such as ours the interior and exterior sectors are coupled and all
allowed modes are modes satisfying boundary conditions of the whole double
wedge.
Thus there can be only one azimuthally constant mode for all $\theta$ (not one
for each sector as one obtains for a perfectly conducting wedge-sheet) hence
the zero mode should be counted once. This is exactly what is done in
Eq.~(\ref{fullEnergy}) because the dispersion function (\ref{DEfix})
has a double
zero at $\nu=0$ cancelling the factor $1/2$. Hence no additional correction
term is necessary and the use of dispersion relations with the argument
principle automatically gives the full result.

In our numerical considerations reported in section \ref{sec_numerics} we find
correspondence with the finite part of the energy reported in \cite{brevik09}
when applied to two complementary wedges separated by a perfectly conducting
sheet. Note how this correspondance is somewhat peculiar: In the energy
expression of that reference the zero mode was counted with half weight where
it should have been accounted for fully, but in adding the energy of the
complementary wedge as in Eq.~(\ref{esum}) each of the complementary wedges
contribute a half of the $m=0$ mode energy, amounting to the full energy when
we insist that this mode be common to the whole system.

It is thus made clear how the divergent term found in
\cite{nesterenko01,nesterenko03,brevik09}
can be seen as a pathology of the ideal conductor boundary conditions at
$\theta=\pm\alpha/2$ which (a) completely removes the azimuthally constant TM
mode and (b) cleanly severs the connection between the interior and exterior of
the wedge. 
Whether a similar term would appear -- perhaps with a finite value -- for a
non-diaphanous wedge remains an open question, since the diaphanous condition
employed herein is also a special case.

\section{Numerical investigation}\label{sec_numerics}

It is useful to introduce the shorthand notation
\be\label{numapprox}
  \tmE = -\frac{1}{4\pi^2 na^2 }\int_0^\infty d\eta \left[\frac{d}{d\eta}
\ln\frac{D}{D_0}\right] Y(\eta),
\ee
where $Y$ is the imaginary part of the integral over the logarithm in
Eq.~(\ref{BigEnergy}),
\be
  Y(\eta) = \int_0^\infty dx x\arctan\frac{- x^2 \im\{\lambda^2_{i\eta}(x)\}}
{1- x^2 \re\{\lambda^2_{i\eta}(x)\}},
\ee
where we take the argument of the logarithm to lie in
$[-\frac{\pi}{2},\frac{\pi}{2}]$.

Near $x=0$ this integrand behaves like $x\sin(\ln x)$, oscillating
increasingly fast. Techniques of rotating the integration path are restricted
by the scarcity of methods for evaluating Bessel functions of general complex
order, and will anyway come at the cost of making $\frac{d}{d\eta}\ln D/D_0$
oscillatory.
For numerical purposes it is more useful to perform the substitution $x=e^y$:
\be
  Y(\eta) = \int_{-\infty}^\infty dy e^{2y} \arctan\frac{- e^{2y}
\im\{\lambda^2_{i\eta}(e^y)\}}{1- e^{2y} \re\{\lambda^2_{i\eta}(e^y)\}}.
\ee
For moderate values of $\eta$ this integrand is numerically manageable (there
are $\mathcal{O}(4\eta)$ significant oscillations to integrate over), the
remaining challenge being the evaluation of $\lambda_{i\eta}(x)$.

\begin{figure}[tb]
  \begin{center}
  \includegraphics[width=2.4in]{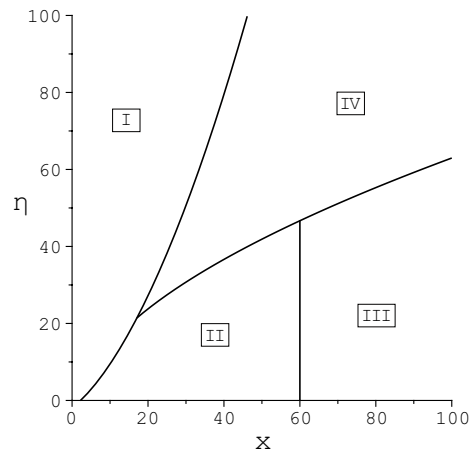} \\
  \includegraphics[width=2.4in]{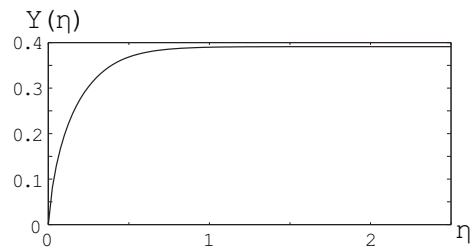}
  \caption{(a)Different methods of calculation used in different areas of the
$x,\eta$ plane (see text). (b) The function $Y(\eta)$.}
  \label{fig_Kie}
  \end{center}
\end{figure}

Rather than consider the complex function $I_{i\eta}(x)$ it is numerically
useful to consider the real function
\be
  L_{i\eta}(x) = \frac1{2}[I_{i\eta}(x)+I_{-i\eta}(x)].
\ee
When $\eta$ is real, $L_{i\eta}(x)=\re I_{i\eta}(x)$. We find, using the
Wronskian relation
\be\label{wronskian}
  \mathcal{W}[K_\nu, I_\nu](x)=1/x
\ee
and relations between the two modified Bessel functions,
that $\lambda_{i\eta}$ can be written as
\be
  \lambda_{i\eta}(x) = \frac1{x} + 2K_{i\eta}'(x)L_{i\eta}(x)
- \frac{2i\sinh \eta\pi}{\pi}K_{i\eta}'(x)K_{i\eta}(x)
\ee

For obtaining the right limit of the integrand near $\eta=0$ one may notice
that $Y(\eta)$ for small $\eta$ is
\[
  Y(\eta) \sim -\eta\int_0^\infty dxx^2\frac{4 K_0K_1(1-2xI_0K_1)}
{x^2- (1-2xI_0K_1)^2} + \mathcal{O}(\eta^2)
\]
plus higher orders. Numerically one finds
\be
  Y(\eta)\sim 0.87442\eta+\mathcal{O}(\eta^2).
\ee

A complete algorithm for evaluating $K$ and $L$ for imaginary order and
real argument was developed by Temme, Gil and Segura \cite{gil04,gil04a}.
Since we
are only calculating products of Bessel functions and the methods for
calculating one is much like that for another, the code performance could be
increased significantly by reprogramming (we used programming language C\#).

Different calculation methods are appropriate in different areas of the
$x,\eta$ plane as shown in Fig.~\ref{fig_Kie}a. For $K$ and $K'$ we use
Maclaurin type series expansion in region I in the figure (bounded by
$\eta > 0.044(x-3.1)^{1.9} +x-3.1$) and in regions II and III (bounded by
$\eta<380(\frac{x-3}{2300})^{0.572}$) a method of continued fractions is
used \cite{gil02} (the continued fractions method of \cite{BookPress92} may
be used for imaginary orders also). No continued fractions method is
available for $L$, but series expansions turn out to be more robust than for
$K,K'$; for $x<60$ (region II) Maclaurin series expansion is used, and
asymptotic series expansion is used above this (region III). In the remaining
area (region IV) Airy function type asymptotic expansions were used
\cite{gil04,temme97,gil03}. In addition a fast method for evaluating complex
gamma functions was necessary - we used that of Spouge \cite{spouge94}.
The resulting algorithm was able to calculate $\lambda_{i\eta}(x)$ with at
least eight significant digits on $x,\eta\in [0,100]$, more than sufficient
for our purposes.

Because the calculation of $\lambda$ is rather elaborate we do not do the
double integral (\ref{numapprox}) directly, but calculate a number of
discrete values of $Y(\eta)$ and use spline interpolation to represent $Y$ in
the integration over $\eta$, which then converges rapidly. The function
$Y(\eta)$ is zero at $\eta=0$ and increases smoothly thence to approach a
positive constant, obtained already at modest values of $\eta$, as plotted in
Fig.~\ref{fig_Kie}b. The factor $[\ln D/D_0]'$ behaves as $e^{-2\eta\alpha}$
for large $\eta$ (assuming $\alpha<\pi$) assuring rapid convergence when
$\alpha$ is not close to zero or $2\pi$.

In the limit $r^2\to 1$ we should obtain correspondence with \cite{brevik09}
where the energy of a perfectly reflecting wedge closed by a diaphanous arc
was considered. In this strong coupling case (the arc becoming perfectly
reflecting) the energy of the sector inside the wedge only (modulo a singular
term) was written on the form
\be\label{previousEp}
  \tmE_\text{id} = \frac1{8\pi na^2}e(p),
\ee
where the dimensionless function $e(p)$ is given in \cite{brevik09}, Eq.\
(4.22), and $p=\pi/\alpha$ as before. In the present case the modes in the
interior and exterior sectors never decouple, even in the limit $r\to 1$ and
Eq.~(\ref{BigEnergy}) thus calculates the energy of the whole system,
regularised by subtracting the energy of free space, that is, by subtracting
the result when the arc is moved to infinity (this is already implicitly
subtracted by use of Eq.~(\ref{circle})) and the wedge boundaries become
transparent. The energy to compare with is therefore on the form given in
\cite{brevik09} where the energy of the complementary wedge is added and that
of a cylinder subtracted. We can therefore write Eq.~(\ref{E_perfMod}) in the
form  $\tmE_\text{id}=\tilde{e}_\text{id}(p)/8\pi na^2$ where
\be\label{esum}
  \tilde{e}_\text{id}(p) = e(p) + e(p') - 2e(1),
\ee
and $p'$ was defined in (\ref{pprime}).

\begin{figure}[tb]
  \begin{center}
  \includegraphics[width=3.4in]{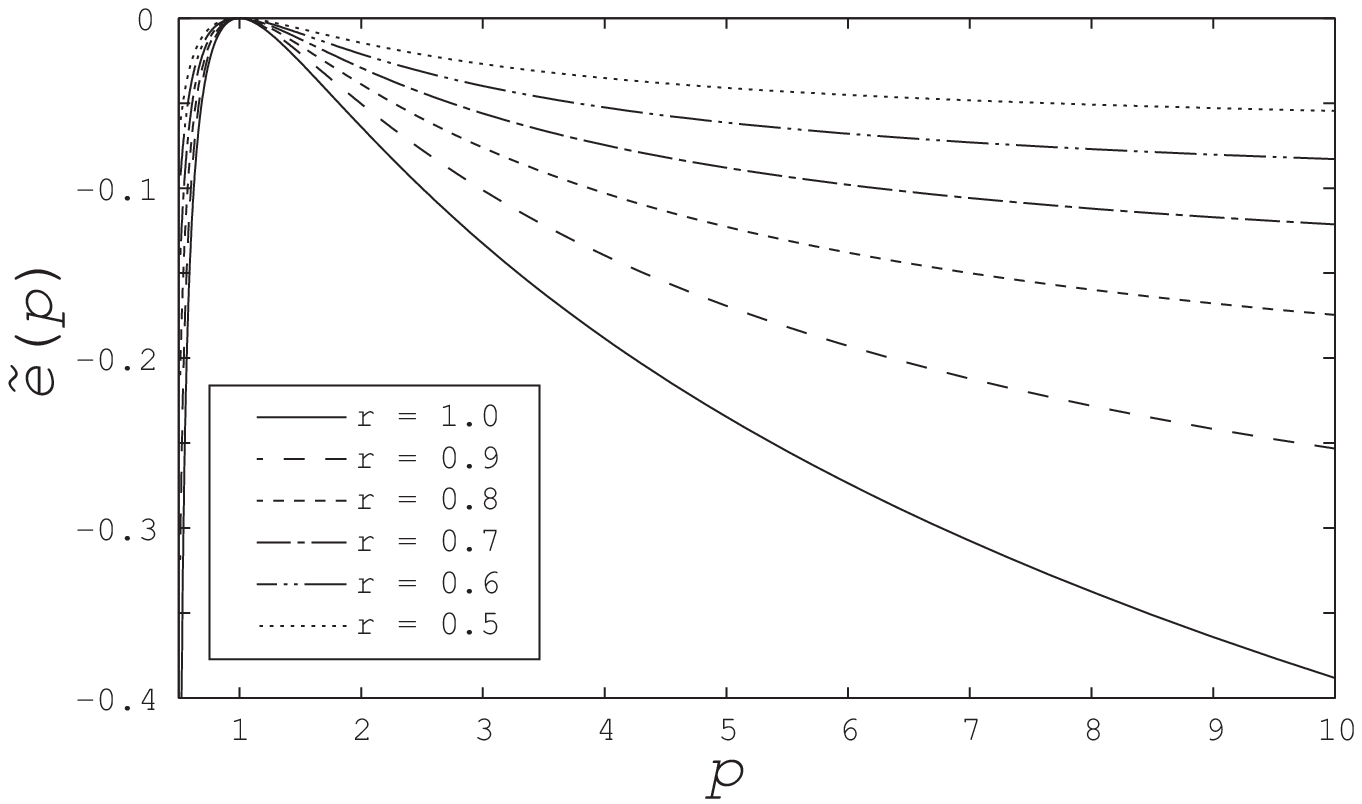} \\
  \includegraphics[width=3.4in]{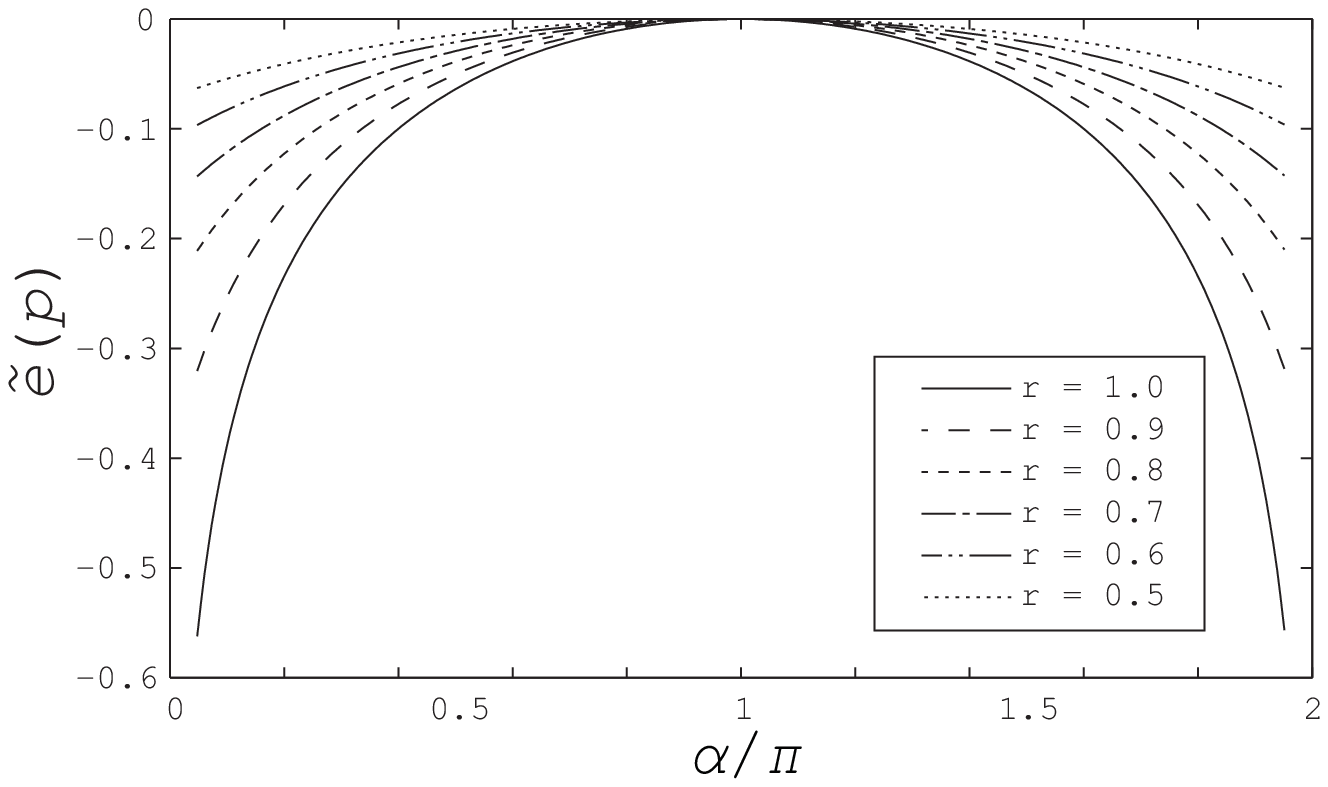}
  \caption{Above: The function $\tilde{e}(p)$ calculated for different $r$.
Below: Same quantity, now plotted as a function of opening angle $\alpha$. }
  \label{fig_ep}
  \end{center}
\end{figure}

For our system the corresponding function is
\be
  \tilde{e}(p) = -\frac2{\pi}\int_0^\infty d\eta \left[\ln\frac{D}{D_0}
\right]^\prime Y(i\eta;r).
\ee
We plot $\tilde{e}(p)$ as a function of $p$ and as a function of $\alpha$ in
Fig.~\ref{fig_ep}. When $r\to 1$ full agreement with
$\tilde{e}_\text{id}(p)$ of Eq.~(\ref{esum}) is obtained.

\section*{Conclusions}

We have analysed the Casimir energy of a magnetodielectric cylinder whose
cross section is a wedge closed by a circular arc under the restriction that
the cylinder be diaphanous, i.e.\ that the speed of light be spatially
uniform. We obtain an expression for the Casimir energy per unit length of the
cylinder, regularised by subtraction of the energy of the wedge alone and the
cylinder alone. The energy is then zero when the opening angle of the wedge,
$\alpha$, equals $\pi$, it is symmetrical under the substitution
$\alpha\leftrightarrow 2\pi-\alpha$, and it remains finite as $\alpha$ tends
to zero
or $2\pi$ except when the absolute reflection coefficients of the wedge
boundaries equal unity.

A numerical investigation confirms that this generalises a previously known
result for a perfectly conducting wedge closed by a diaphanous
magnetodielectric arc in the limit where the arc becomes perfectly reflecting,
except for a singular term present in that geometry which we argue does not
present itself in the present configuration. This implies that the singular
term found and discussed in \cite{brevik09} is an artefact of the use
of ideal conductor boundary conditions and does not enter for a diaphanous wedge.

We mention finally that the diaphanous condition
$\epsilon\mu=$ constant is an important simplifying element in the
analysis. If this condition were given up, the problem would be
very difficult to solve. As mentioned also in \cite{brevik09}, the
effect is the same as that encountered in the Casimir theory  of a
solid ball: the condition of diaphanousness causes the divergent
terms to {\it vanish} \cite{brevik82}. Analogously, when
calculating the Casimir energy for a piecewise uniform string, the
same effect turns up. If the velocity of sound (in this case sound
replaces light) is the same $(=c)$ in in the different pieces of
the string, then the theory works smoothly \cite{brevik90}. If
this condition is relaxed, the problem becomes in practice
intractable.

\acknowledgments
The work of KAM was supported in part by grants from the US National Science
Foundation and the US Department of Energy. S\AA E thanks Carsten Henkel and
Francesco Intravaia for stimulating discussions on this topic.
KAM thanks Klaus Kirsten, Prachi Parashar, and Jef Wagner for collaboration.
The authors are grateful to Vladimir Nesterenko for useful remarks on the
manuscript.

\appendix
\section{Semitransparent wedge}
In this appendix we sketch another way of deriving the azimuthal
dependence, based on an analogous scalar model, in which the
wedge is described by a $\delta$-function potential,
\be
V(\theta)=\lambda_1\delta(\theta-\alpha/2)+\lambda_2\delta(\theta+\alpha/2).
\label{dfpot}
\ee
This has the diaphanous property of preserving the speed of
light both within and outside the wedge.  We can solve this cylindrical
problem in terms of the two-dimensional Green's function $G$, which
satisfies
\bea&&\left[-\frac1\rho\frac\partial{\partial \rho}\rho
\frac\partial{\partial\rho}
+\kappa^2-\frac1{\rho^2}\frac{\partial^2}{\partial\theta^2}
+\frac{V(\theta)}{\rho^2}
\right]G(\rho,\theta;\rho',\theta')\nonumber\\
&&\qquad=\frac1\rho\delta(\rho-\rho')\delta(\theta-\theta').
\eea
This separates into two equations, one for the angular eigenfunction
$\Theta_\nu(\theta)$
\be
\left[-\frac{\partial^2}{\partial\theta^2}+V(\theta)\right]\Theta_\nu(\theta)
=\nu^2\Theta_\nu(\theta),\label{eveq}
\ee
leaving us with the radial reduced Green's function equation,
\be
\left[-\frac1\rho\frac\partial{\partial \rho}\rho\frac\partial{\partial \rho}
+\kappa^2
+\frac{\nu^2}{\rho^2}\right]g(\rho,\rho')=\frac1\rho\delta(\rho-\rho').
\ee
The latter, for a Dirichlet arc at $\rho=a$, has the familiar solution,
\begin{subequations}
\bea
g(\rho,\rho')&=&I_\nu(\kappa \rho_<)K_\nu(\kappa \rho_>)-I_\nu(\kappa \rho)
I_\nu(\kappa \rho')
\frac{K_\nu(\kappa a)}{I_\nu(\kappa a)},\nonumber\\
&&\qquad\qquad\quad \rho,\rho'<a,\\
g(\rho,\rho')&=&I_\nu(\kappa \rho_<)K_\nu(\kappa \rho_>)-K_\nu(\kappa \rho)
K_\nu(\kappa \rho')
\frac{I_\nu(\kappa a)}{K_\nu(\kappa a)},\nonumber\\
&&\qquad\qquad\quad \rho,\rho'>a.
\eea
\end{subequations}

The azimuthal eigenvalue $\nu$ is determined by Eq.~(\ref{eveq}).  For the
wedge $\delta$-function potential (\ref{dfpot}) it is easy to determine
$\nu$ by writing the solutions to Eq.~(\ref{eveq})
as linear combinations of $e^{\pm i\nu\theta}$,
with different coefficients in the sectors $|\theta|<\alpha/2$ and $\pi\ge
|\theta|>\alpha/2$. Continuity of the function, and discontinuity of
its derivative, are imposed at the wedge boundaries. The four simultaneous
linear homogeneous equations have a solution only if the secular equation
is satisfied:
\bea
0=D(\nu)
&=&\sin^2\nu(\alpha-\pi)-\left(1-\frac{4\nu^2}{\lambda_1\lambda_2}\right)
\sin^2\pi\nu
\nonumber\\
&&\quad\mbox{}-\left(\frac{\nu}{\lambda_1}+\frac\nu{\lambda_2}\right)
\sin2\pi\nu.\label{a6}
\eea
Because we recognize that the reflection coefficient for a single
$\delta$-function interface is
$r_i=(1+2i\nu/\lambda_i)^{-1}$, so
\be
\re r_1^{-1}r_2^{-1}=1-\frac{4\nu^2}{\lambda_1\lambda_2},\quad
\im r_1^{-1}r_2^{-1}=\frac{2\nu}{\lambda_1}+\frac{2\nu}{\lambda_2},\label{a7}
\ee
we see that this dispersion relation coincides with that
in Eq.~(\ref{DEfix}) when the reflection coefficient is purely real.
Note that the $\nu=0$ root of Eq.~(\ref{a6}) is spurious and must be excluded;
there are no $\nu=0$ modes for the semitransparent wedge.

Now the full Green's function can be constructed as
\bea
G(x,x')&=&\int\frac{d\omega}{2\pi}e^{-i\omega(t-t')}\int\frac{dk}{2\pi}
e^{ik(z-z')}\nonumber\\
&&\qquad\times\frac1{2\pi}\sum_\nu
\Theta_\nu(\theta)\Theta_\nu^*(\theta')g_\nu(\rho,\rho'),
\eea
from which the Casimir energy per length can be computed from
\be
\mathcal{E}=\frac1{2i}\int_{-\infty}^\infty \frac{d\omega}{2\pi}2\omega^2
\int_{-\infty}^\infty \frac{dk}{2\pi}\sum_\nu \int_0^\infty d\rho\,\rho\,
g_\nu(\rho,\rho),
\ee
where we have recognized that
because the eigenvalue equation for $\nu$ is a Sturm-Liuoville problem,
the integration over the $\theta$ eigenfunctions is $2\pi$.  As above,
we can enforce the eigenvalue condition by the argument principle,
so we have the expression after again converting to polar coordinates
as in Eq.~(\ref{pcoord}),
\be
\mathcal{E}=\frac1{8\pi^2i}\int_0^\infty d\kappa\,\kappa^3\int_{-\infty}
^\infty d\eta\left(\frac{d}{d\eta}\ln D(i\eta)\right)\int_0^\infty d\rho\, \rho \,
g_{i\eta}(\rho,\rho).
\ee
Further,  we must
subtract off the free radial Green's function without the arc at $r=a$,
which then implies
\be
\int_0^\infty d\rho\,\rho\,g_{i\eta}(\rho,\rho)\to\frac{a}{2\kappa}
\frac{d}{d\kappa a}
\ln[I_{i\eta}(\kappa a)K_{i\eta}(\kappa a)],
\ee
as well as remove the term present without the wedge potential:
\be
D(\nu)\to\tilde D(\nu)=\frac{\lambda_1\lambda_2}{4\nu^2}\frac{D(\nu)}{\sin^2\pi\nu},
\ee
leaving us with an expression for the Casimir energy analogous to
Eq.~(\ref{BigEnergy}).  This can be further simplified by noting that
$\frac{d}{d\eta}\ln \tilde D(i\eta)$ is odd, which then yields the
expression
\be
\mathcal{E}=-\frac1{4\pi^2 a^2}\int_0^\infty dx\,x\int_0^\infty d\eta
\left(\frac{d}{d\eta}\ln\tilde D(i\eta)\right)
\arctan\frac{K_{i\eta}(x)}{L_{i\eta}(x)},
\ee
where
\begin{subequations}
\bea
K_\mu(x)&=&-\frac\pi{2\sin\pi\mu}\left[I_\mu(x)-I_{-\mu}(x)\right],\\
L_\mu(x)&=&\frac{i\pi}{2\sin\pi\mu}\left[I_\mu(x)+I_{-\mu}(x)\right],
\eea
\end{subequations}
where both $L_{i\eta}(x)$ and $K_{i\eta}(x)$ are real for real $\eta$ and
$x$, and
\be
I_{i\eta}(x)=\frac{\sinh\eta\pi}\pi[L_{i\eta}(x)-iK_{i\eta}(x)].
\ee

Details of the calculation of the Casimir energy for a semitransparent
wedge will appear elsewhere.



\begin{thebibliography}{99}
\bibitem{casimir48} H. B. G. Casimir, Proc. Kon. Ned. Akad. Wetensch. {\bf 51}, 793 (1948).
\bibitem{lifshitz55} E. M. Lifshitz, Zh. Eksp. Teor. Fiz. {\bf 29}, 94 (1955) [Sov. Phys. JETP {\bf 2}, 73 (1956)]

\bibitem{BookMilton01} K. A. Milton {\it The Casimir Effect: Physical Manifestations of the Zero-Point Energy} (World Scientific, Singapore, 2001)
\bibitem{milton04} K. A. Milton, J. Phys. A: Math. Gen. {\bf 37}, R209 (2004)
\bibitem{lamoreaux05} S. K. Lamoreaux, Rep. Prog. Phys. {\bf 68}, 201 (2005)
\bibitem{buhmann07} S. Y. Buhmann and D.-G. Welsch, Prog. Quantum Electron. {\bf 31}, 51 (2007)

\bibitem{deraad81} L. L. DeRaad, Jr.\ and K. A. Milton, Ann. Phys. {\bf 186}, 229 (1981)
\bibitem{brevik94} I. Brevik and G. H. Nyland, Ann. Phys. {\bf 230}, 321 (1994)
\bibitem{gosdzinsky98} P. Gosdzinsky and A. Romeo, Phys. Lett. B {\bf 441}, 265 (1998)
\bibitem{milton99} K. A. Milton, A. V. Nesterenko and V. V. Nesterenko, Phys. Rev. D {\bf 59}, 105009 (1999)
\bibitem{lambiase99} G. Lambiase, V. V. Nesterenko and M. Bordag, J. Math. Phys. {\bf 40}, 6254 (1999)
\bibitem{caveroPelaez05} I. Cavero-Pel\'{a}ez and K. A. Milton, Ann. Phys. {\bf 320}, 108 (2005); J. Phys. A. {\bf 39}, 6225 (2006)
\bibitem{romeo05} A. Romeo and K. A. Milton, Phys. Lett. B {\bf 621}, 309 (2005); J. Phys. A {\bf 39}, 6703 (2006)
\bibitem{brevik07} I. Brevik and A. Romeo, Phys. Scripta {\bf 76}, 48 (2007)
\bibitem{caveroPelaez06} I.~Cavero-Pel\'{a}ez, K.~A.~Milton and K.~Kirsten, J.\ Phys.\ A  {\bf 40}, 3607 (2007)

\bibitem{dowker78} J. S. Dowker and G. Kennedy, J. Phys. A {\bf 11}, 895 (1978)
\bibitem{deutsch79} D. Deutsch and P. Candelas, Phys. Rev. D {\bf 20}, 3063 (1979)
\bibitem{brevik96} I. Brevik and M. Lygren, Ann. Phys. {\bf 251}, 157 (1996)
\bibitem{brevik98} I. Brevik, M. Lygren and V. Marachevsky, Ann. Phys. {\bf 267}, 134 (1998)
\bibitem{brevik01} I. Brevik, K. Pettersen, Ann. Phys. {\bf 291}, 267 (2001)
\bibitem{nesterenko02} V. V. Nesterenko, G. Lambiase and G. Scarpetta, Ann. Phys. {\bf 298}, 403 (2002)
\bibitem{razmi05} H. Razmi, S. M. Modarresi, Int. J. Theor. Phys. {\bf 44}, 229 (2005).
\bibitem{BookMostepanenko97} V. M. Mostepanenko and N. N. Trunov, {\it The Casimir Effect and
Its Applications} (Oxford University Press, Oxford, 1997).
\bibitem{nesterenko01} V. V. Nesterenko, G. Lambiase and G. Scarpetta, J. Math. Phys. {\bf 42}, 1974 (2001)
\bibitem{nesterenko03} V. V. Nesterenko, I. G. Pirozhenko and J. Dittrich, Class. Quantum Grav. {\bf 20}, 431 (2003)
\bibitem{rezaeian02} A. H. Rezaeian and A. A. Saharian, Clas. Quant. Grav. {\bf 19}, 3625 (2002)
\bibitem{saharian07} A. A. Saharian, Eur. Phys. J. C {\bf 52}, 721 (2007)
\bibitem{saharian09} A. A. Saharian, in {\it The Casimir Effect and Cosmology: A volume in honour of Professor Iver H. Brevik on the occasion of his 70th birthday} S. Odintsov et al. (eds.) (Tomsk State Pedagogical University Press, 2008), p.87, preprint {\it hep-th/0810.5207} .
\bibitem{mendes08} T. N. C. Mendes, F. S. S. Rosa, A. Ten\'{o}rio and C. Farina, J. Phys. A {\bf 41} 164029 (2008)
\bibitem{rosa08} F. S. S. Rosa, T. N. C. Mendes, A. Ten\'{o}rio and C. Farina, Phys. Rev. A {\bf 78} 012105 (2008)
\bibitem{barton87} G. Barton, Proc. R. Soc. London {\bf 410}, 175 (1987)
\bibitem{skipsey05} S. C. Skipsey, G. Juzeli\={u}nas, M. Al-Amri, and M. Babiker, Optics Commun. {\bf 254}, 262 (2005)
\bibitem{skipsey06} S. C. Skipsey, M. Al-Amri, M. Babiker, and 
G. Juzeli\={u}nas, Phys. Rev. A {\bf 73}, 011803(R) (2006)

\bibitem{sukenik93} C. I. Sukenik, M. G. Boshier, D. Cho, V. Sandoghdar and E. A. Hinds, Phys. Rev. Lett. {\bf 70} 560 (1993)

\bibitem{brevik09} I. Brevik, S. \AA. Ellingsen and K. A. Milton, Phys. Rev. E {\bf 79} 041120 (2009)


\bibitem{macdonald1895} H. M. Macdonald, Proc. Lond. Math. Soc. {\bf 26}, 156 (1895)
\bibitem{sommerfeld1896} A. Sommerfeld, Math. Ann. {\bf 47} 317 (1896)
\bibitem{malyuzhinetsThesis50} G. D. Malyuzhinets, Ph.D. Thesis, P. N. Lebedev Phys. Inst. Acad. Sci. USSR (1950)
\bibitem{osipov99} A. V. Osipov and A. N. Norris, Wave Motion {\bf 29}, 313 (1999)
\bibitem{osipov98} A. V. Osipov and K. Hongo, Electromagnetics {\bf 18}, 135 (1998)
\bibitem{kontorovich38} M. J. Kontorovich and N. N. Lebedev, Zh. Exp. Theor. Fiz {\bf 8}, 1192 (1938)
\bibitem{oberhettinger54} F. Oberhettinger, Commun. Pure \& App. Math. {\bf 7}, 551 (1954)
\bibitem{osipov93} A. V. Osipov, Probl. Diff. Prop. Waves {\bf 25} 173 (1993)
\bibitem{knockaert97} L. Knockaert, F. Olyslager and D. De Zutter, IEEE Trans. Antennas Propag. {\bf 45}, 1374 (1997)
\bibitem{rawlins99} A. D. Rawlins, Proc. R. Soc. Lond. A {\bf 455}, 2655 (1999)
\bibitem{salem06} M. A. Salem, A. H. Kamel and A. V. Osipov, Proc. R. Soc. A {\bf 462}, 2503 (2006).
\bibitem{salem08} M. A. Salem and A. H. Kamel, Q. J. Mech. Appl. Math. {\bf 61}, 219 (2008)

\bibitem{vankampen68} N. G. van Kampen, B. R. A. Nijboer and K. Schram K, Phys. Lett. {\bf 26A}, 307 (1968)


\bibitem{BookDitkin65} V. A. Ditkin and A. P. Prudnikov, {\it Integral Transforms and Operational Calculus} (Oxford: Pergamon Press, 1965) Ch.\ 11
\bibitem{BookOberhettinger72} F. Oberhettinger, {\it Tables of Bessel Transforms} (Berlin: Springer, 1972) Ch.\ 5
\bibitem{gautschi06} W. Gautschi, BIT {\bf 46} 21 (2006).

\bibitem{schwinger78} J. Schwinger, L. L. DeRaad, Jr. and K. A. Milton, Ann. Phys. {\bf 115}, 1 (1978)
\bibitem{ellingsen07} S. A. Ellingsen and I. Brevik, J. Phys. A {\bf 40}, 3643 (2007)
\bibitem{tomas95} M. S. Toma\v{s}, Phys. Rev. A {\bf 51}, 2545 (1995)
\bibitem{brevik01b} I. Brevik, B. Jensen, and K. A. Milton, Phys. Rev. D {\bf 64}, 088701 (2001)
\bibitem{nesterenko06} V. V. Nesterenko, J. Phys. A, {\bf 39}, 6609 (2006)
\bibitem{BookStratton41} J. A. Stratton, {\it Electromagnetic Theory} (New York: McGraw Hill, 1941) \S 9.15
\bibitem{BookParsegian06} V. A. Parsegian, {\it Van Der Waals Forces} (Cambridge: Cambridge Univ. Press, 2006) \S L3.3
\bibitem{nesterenko08} V. V. Nesterenko, J. Phys. A, {\bf 41}, 164005 (2008)
\bibitem{ellingsen09} S. \AA. Ellingsen, in {\it The Casimir Effect and 
Cosmology: A volume in honour of Professor Iver H. Brevik on the occasion of 
his 70th birthday} S. Odintsov et al. (eds.) (Tomsk State Pedagogical 
University Press, Tomsk, Russia, 2008), p.45, preprint 
{\it quant-ph/0811.4214}.
\bibitem{ambjorn83} J. Ambj\o rn and S. Wolfram, Ann. Phys. {\bf 147}, 1 (1983)




\bibitem{gil04} A. Gil, J. Segura, N.M. Temme, ACM Trans. Math. Soft. {\bf 30} 145 (2004)
\bibitem{gil04a} A. Gil, J. Segura, N.M. Temme, ACM Trans. Math. Soft. {\bf 30} 159 (2004)
\bibitem{gil02} A. Gil, J. Segura, N.M. Temme, J. Comp. Phys. {\bf 175} 398 (2002)
\bibitem{BookPress92} W. Press et.al., {\it Numerical Recipes} 2nd ed.\ (Cambridge: Cambridge Univ. Press, 1992) \S 6.6
\bibitem{temme97} N.M. Temme, Num. Alg. {\bf 15} 207 (1997)
\bibitem{gil03} A. Gil, J. Segura, N.M. Temme, J. Comp. App. Math. {\bf 153} 225 (2003)
\bibitem{spouge94} J.L. Spouge, SIAM J. Numer. Anal. {\bf 31} 931 (1994)

\bibitem{brevik82}
I. Brevik and H. Kolbenstvedt, Ann. Phys. (N.Y.) {\bf 143}, 179
(1982); {\bf 149}, 237 (1983).
\bibitem{brevik90}
To our knowledge the first paper in this direction was I. Brevik
and H. B. Nielsen, Phys. Rev. D {\bf 41}, 1185 (1990). A survey is
given by I. Brevik, A. A. Bytsenko, and B. M. Pimentel, in {\it
Theoretical Physics 2002 (Horizons in World Physics)}, edited by
T. F. George and H. F. Arnoldus (Nova Science, New York, 2002).
\end{thebibliography}
\end{document}